# Characterization of ultra-low carbon steel: A preliminary approach to investigate the quality and standards of locally-available steel

[1]Sulaiman, [1]Falak Niaz, [1]Muhammad Riaz Khan, [1,2,*]Saeed Ullah, [3]Azim Khan, Abdur Rahim[4] and [3]Syed Sohail Ahmad Shah

[1]Department of Physics, University of Peshawar, Peshawar, KP Pakistan

[2]Instituto de Física, Universidade de São Paulo, Caixa Postal 66318 - CEP 05315-970, São Paulo, SP, Brazil

[3]Institute of Metal Research, Chinese Academy of Sciences, 72 Wenhua Road, Shenyang, Liaoning, 110016, China

[4]Department of Physics, Khushal Khan Khattak University, Karak, Pakistan.

**Summary:** In the present work, experimental study has been carried out to expose the thermal, mechanical, and microstructural properties of low carbon steel as well as to inspects the influence of etchant concentration and etching time on its microstructure. Ultra-low carbon steel, in the form of a sheet, was collected from the Mughal Steel Industry, Peshawar, Pakistan. The sample was chemically etched, using Nital as an etchant, by two different methods: first, by changing the etching time while keeping the composition of etchant the same and second, by keeping the time constant while varying the etchant composition in a range of 5-14 %. The microstructure analysis revealed that ultra-fine grain can be obtained for the etchant composition of 8 % nitric acid in ethanol. Additionally, we noticed that the best etching time, for getting a clear morphology, was 90 s. The X-ray diffraction revealed mainly α-iron. Thermal analysis showed a minor weight loss followed by weight gain of 1.31 wt %. Contraction and expansion, observed on the TDA curve, suggested the transformation of BCC to FCC structure. Our results indicated that the specimen is highly ductile, malleable and soft.

## 1 Introduction

The ever-increasing climate changes have led to the massive demands for engineering new and emergent materials with improved properties [1,2]. For that reason, researchers and engineers are working to manufacture new and advanced materials as well as modifying the properties of the existing materials for the broad range applications. Different types of materials have been engineered for multiple industrial innovations [3-7]. Among them, metal-based materials are considered as a manufacturing backbone of industrialized civilization due to their higher ductility, load-bearing capacity, and damage tolerance [8]. Currently, such materials are the subject of widespread research efforts [9-11]. Advancing the safety standards, in conjunction with degrading energy consumption, is a predominant target in modern mobility concepts. Therefore, the development of structural steel with improved mechanical properties is a key topic in steel research which may facilitate the light-weight construction design for example building, bridges, and automotive structures [9-14].

Carbon steel is one of the key material for transportation, infrastructure, and modern manufacturing [10.11]. It has drawn considerable attention due to the enhanced mechanical properties that make it well-suited for applications in the fields of energy, health, machinery, construction industry, and hydraulic and marine structures. Examples include Fe-Cr steel in turbines [15], Fe-Mn steels for light-weight and passenger protecting mobility [16], and Fe-Cr steels for the nuclear and fusion power plants [17]. Although massive research has been conducted on steel [9-12], however, there has been few or no pertinent information about the production standards and quality of the locally-made steels as compared to the one in the developed countries. Therefore, an attempt is made here to investigate the standards and quality of locally-available steel. We believe that the present investigation may accelerate the commercialization of locally-available steel.

## 2 Materials and Experiments

The material studied in present work was an ultra-low carbon steel collected from the Mughal steel industry, Pakistan. According to the data provided by manufacturer industry, it contains the following alloying elements:

Iron - 99 %, Sulfur - less than 0.04 %, Manganese - 0.35 to 0.65 %, Carbon - 0.17 to 0.24 %, Chromium - up to 0.3 %, Phosphorous - less than 0.04 %, and Nickel - less than 0.3%. The sample was cut into a size of interest using an abrasive cutoff machine and its apparent density (7.46 g/cm³) was calculated from the mass and dimension. The sectioning-created damages were minimized by grinding and further removed by subsequent polish operations. After polishing, the sample was chemically etched by two different methods using Nital (Nitric acid in ethanol) as an etchant. To see which technique give best results the sample was first etched by changing the etching time while keeping the etching composition the same and then etched by varying the etchant composition while keeping etching time constant.

The mechanical properties were assessed using an electromechanical universal testing machine (UTM) which yields the result at the output in the form of a graph obtained by a computer built-in software where further test results can be retrieved from it. To determine the temperatures associated with significant phase transitions, thermal gravimetric and differential thermal analysis (TG/DTA) was carried out, using Perkin Elmer TG/DTA unit, in the temperature range of 30-1200 $^{o}$C with a heating/cooling rate of 10 $^{o}$C/min. The thermal dilatometric analysis (TDA) was used to measure the dimensional changes of the material, including expansion, contraction and Curie point etc, as a function of temperature [18,19]. The phase analysis was performed by a JEOL X-ray diffractometer (model JDX-3532) equipped with Cu K$_\alpha$ radiations operated at 30 mA and 40 kV in the 2θ range of 10-70$^{o}$ with a step size of 0.02$^{o}$. The microstructure was studied using scanning electron microscope SEM (JEOL JSM 5910) while the elemental compositions were detected using energy-dispersive X-ray spectroscopy (EDS).

## 3 Results and Discussions

### 3.1 Mechanical Analysis

The force versus elongation, obtained by employing UTM, is demonstrated in Fig. 1. The specimen was broken when applied force was about 7800 N with elongation of ~ 21.75 mm which confirmed that the material has large elongation and high load-bearing capacity. The curve revealed that the specimen is highly ductile, malleable and soft. The mechanical properties retrieved from the tensile test are summarized in Table 1.

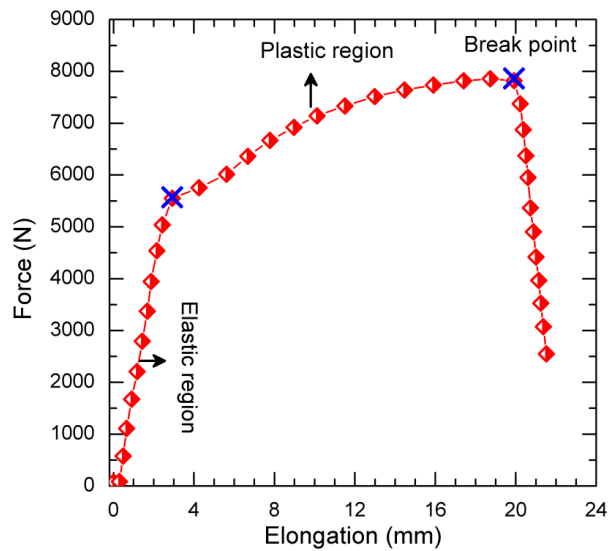

**Fig. 1** Load versus elongation of low carbon steel.

**Table 1** Mechanical properties of ultra-low carbon steel.

| Area (mm$^2$) | Elongation Break (mm) | Load Break (N) | Young's Modulus (N/mm$^2$) | Elongation Peak(mm) | Energy Break(Nm) | Load Peak(N) | Strain Break(%) |
|---|---|---|---|---|---|---|---|
| 19.0 | 21.75 | 2526 | 10489 | 19.72 | 138.43 | 7859 | 17.931 |

### 3.2 Thermal and Phase Analysis

#### 3.2.1 TG/DTA Analysis

TG/DTA analysis was carried out to see the variation in physical properties of the material with rising temperature [20,21]. Fig. 2(a) shows the TG/DTA curve of the specimen measured in the temperature range from room temperature to 1200 °C. The observed minor wt % loss (~0.73 wt %) in the temperature range of 30-785 °C may be due to the removal of entrapped surficial water or other oxides. Furthermore, the sample gained an extra weight (1.31 wt %) with a gain rate of about $3.2 \times 10^{-3}$ mg/K in the temperature range of 785-1200 °C which may be due to the chemical reaction of nitrogen, used as a purging gas, forming nitrides. The DTA curve showed a downward slope with increasing temperature up to 500 °C indicating the endothermic reaction in the sample. According to the phase diagram [22] FeC$_3$ cementite present in the sample is dissolved in ferrite iron which is a BCC crystal having magnetic properties up to ~ 770 °C. An exothermic behavior was observed at a temperature above 500 °C where a small peak was found at 727 °C suggesting that above this temperature the ferrite iron converts into two-phase (α + γ) iron but still retaining its BCC crystal structure. Additionally, a rapid increase was also observed at 910 °C which revealed that beyond this temperature the (α + γ) iron undergoes a phase transition from two phase (BCC) structural iron to single phase (FCC) γ-iron, called austenite [22,23].

#### 3.2.2 TDA Result

The TDA analysis was carried out to determines both reversible and irreversible changes in the length expansion and shrinkage during heating/cooling process [24]. The obtained differential Coefficient of expansion was plotted as a function of sample temperature [see Fig. 2(b)]. We didn't observe any dependence up to 600 °C, however, beyond this temperature, the curve showed a contraction of specimen till 723 °C. The observed contraction may be associated with the magnetic alignments in the sample up to

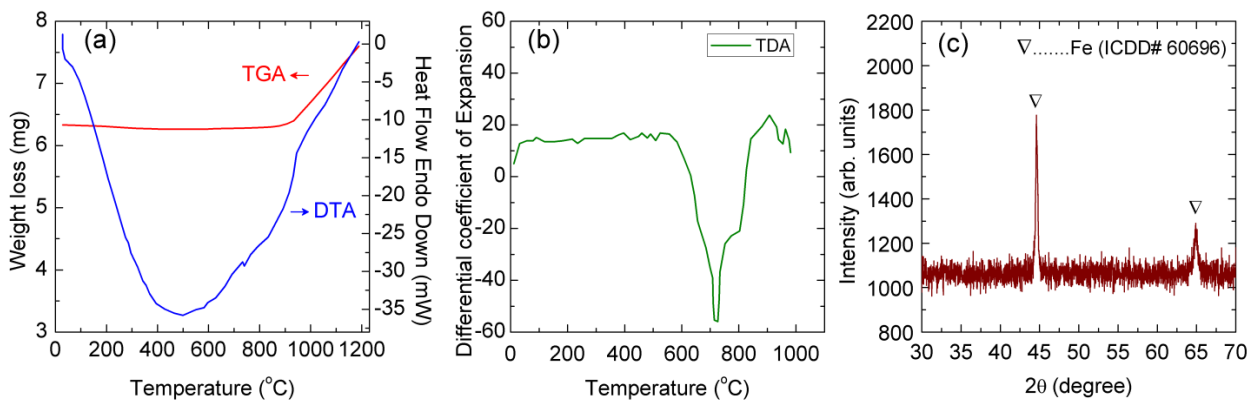

**Fig. 2** (a) TG/DTA curve of the sample. (b) DTA curve of the specimen. (c) The room temperature X-ray diffraction pattern of the sample obtained for 2θ from 30 to 70°.

770 °C (Curie temperature). Further increase in temperature leads to the sample expansion, however, beyond 910 °C the sample start contract again possibly due to the transformation of BCC structure to FCC structure (γ-iron) [23].

### 3.2.3 Phase Analysis

The X-ray diffraction (XRD) pattern of the sample, observed at room temperature in the range of $2\theta = 30\text{-}70°$, is shown in Fig 2(c). To index the observed peaks, the international center for diffraction data (ICDD) was taken into consideration. The d-spacing and relative intensity corresponding to observed XRD peaks (at $2\theta = 44.59°, 65°$) matched the ICDD card# 60696 for **α**-iron.

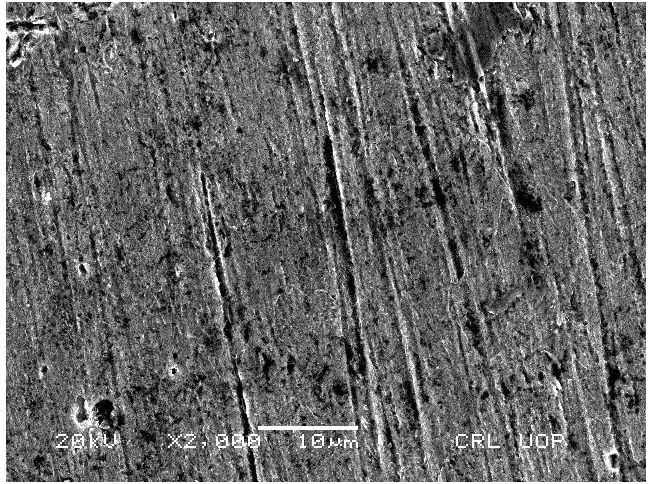

**Fig. 3** Scanning electron micrograph of an un-etched, roughly-polished sample.

### 3.3 Microstructural Analysis

The microstructural analysis was carried out for the specimen etched by either changing the etchant concentration or etching time. To obtain a clear morphology, the sample must be well-polished and chemically etched. It is difficult to get beneficial information without etching, see for example, a scanning electron micrograph of an un-etched, poorly polished sample showing the cutting marks formed during cutting of sample (Fig. 3).

### 3.3.1 Morphology of Sample by Changing Time of Etching

The secondary electron SEM images of the samples etched by keeping the composition of etchant constant while changing the etching time are shown in Fig. 4. The etchant used was 3% nitric acid in ethanol. A clear morphology was obtained when the sample was etched for 1 minute and 30 seconds. The microstructure revealed grains with a grain size of about 208.81 $\mu m^2$. The elemental composition was obtained employing energy dispersive X-ray spectroscopy (EDS). EDS analysis detected mainly iron as a major component together with small amount of carbon, aluminum, manganese and oxygen as impurities. The EDS-detected compositional analysis, summarized in Table 2, is consistent with the XRD finding (sec. 3.2.3).

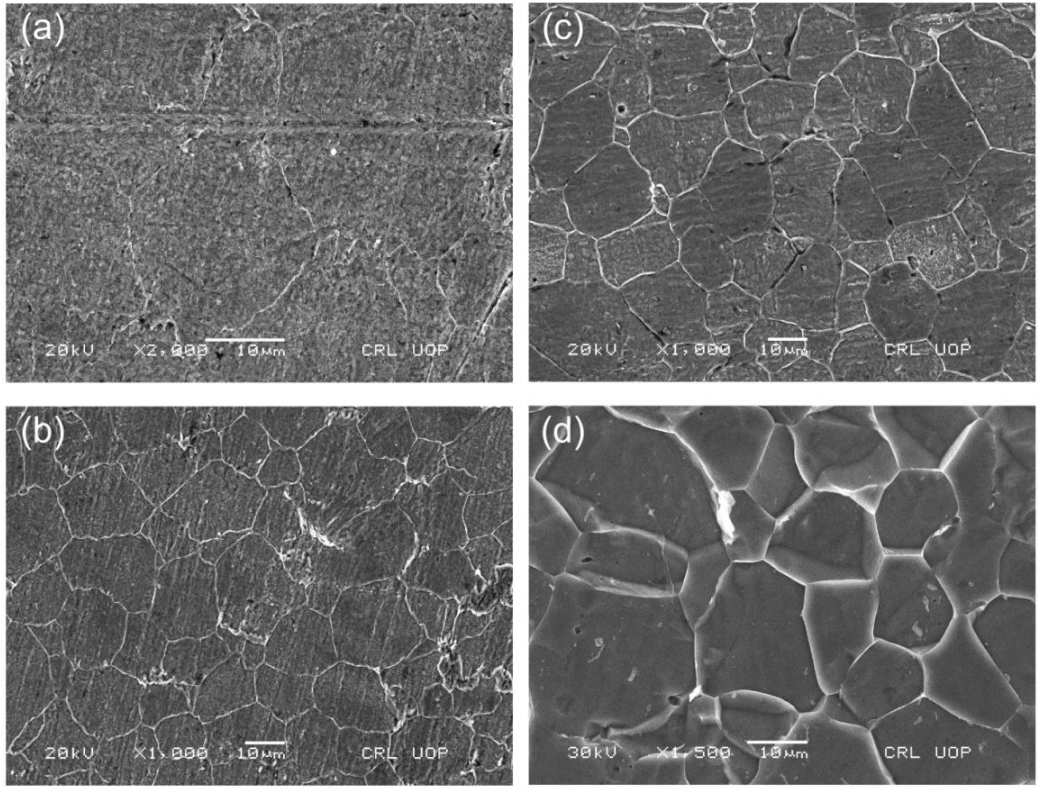

**Fig. 4** SEM micrographs of the specimen by keeping the etchant concentration constant ( 3% nitric acid in ethanol) while changing the time of etching (a) 25 sec (b) 45 sec (c) 60 sec and (d) 90 sec.

**Table 2** EDS-detected compositions in the sample.

| Elements | Wt % | Atomic % |
|---|---|---|
| Fe | 96.63 | 89.75 |
| C | 0.76 | 3.26 |
| Mn | 0.42 | 0.40 |
| Al | 0.39 | 0.74 |
| O | 1.80 | 5.85 |
| Total | 100 | 100 |

**3.2.2 Morphology of Sample by changing concentration of Etching**

Fig. 5 shows the secondary electron SEM images taken after etching the sample by changing the etchant composition while keeping the time of etching same (60 sec). The microstructures revealed that ultra-fine grains can be obtained for the sample etched with 8 % nitric acid in ethanol. Additionally, we found that the sample surface was damaged or eaten by etchant when the concentration was further increased up to 14 %.

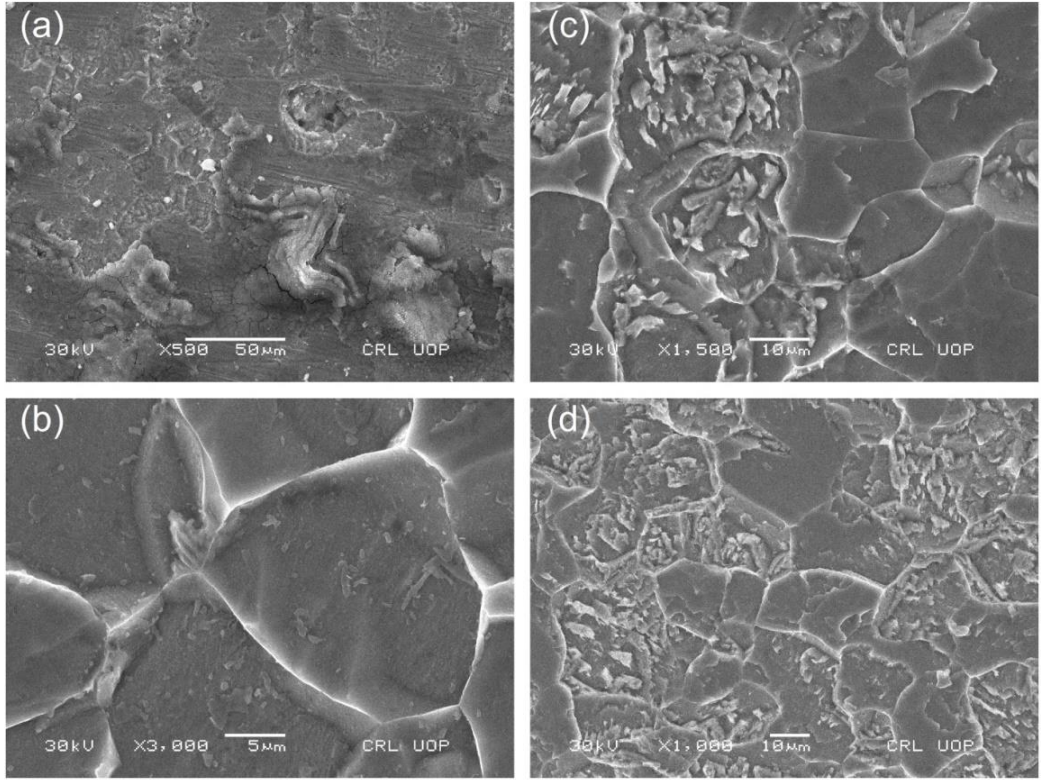

**Fig. 5** SEM images of the samples etched for 60 seconds while changing the etchant concentration (a) 5 % nitric acid in ethanol, (b) 8 % nitric acid in ethanol, (c) 12 % nitric acid in ethanol, and (d) 14 % nitric acid in ethanol.

**4 Conclusions**

In conclusion, we have studied the thermal, mechanical, and microstructural properties of ultra-low carbon steel. We found that the specimen is highly ductile, malleable and soft. TG/DTA analysis revealed an overall weight gain of 1.31 wt % in the temperature range of 785-1200 $^{o}$C. The contraction and expansion, observed on TDA curve, showed a phase transition from dual phase BCC structural iron to single phase FCC austenite. Phase analysis detected only a single phase of α-Ferritic. The variation of etchant concentration resulted in fine grains for 8 % nitric acid in ethanol while changing the etchant time best result was obtained for $t = 90$ s. We believe that our findings may reduce the confusion of the users for the selection of Mughal steel.

**Acknowledgments**
The authors greatly acknowledge the research facility at the Department of Physics, University of Peshawar, Pakistan.